\documentclass{vgtc}                          % final (conference style)
\graphicspath{{figures/}{pictures/}{images/}{./}} % where to search for the images

\usepackage{times}                     % we use Times as the main font
\usepackage{tabu}                      % only used for the table example
\usepackage{booktabs}                  % only used for the table example
\usepackage{lipsum}                    % used to generate placeholder text
\usepackage{mwe}                       % used to generate placeholder figures

\usepackage{mathptmx}                  % use matching math font

\onlineid{0}

\vgtccategory{Research}

\vgtcinsertpkg

\title{Evaluating a Visual Query
Tracer and Builder for Learning~Declarative~Logic Programming}

\author{Juli\'{a}n M\'{e}ndez\thanks{e-mail: julian.mendez2@tu-dresden.de}\\ %
    \parbox{1.8in}{\scriptsize \centering Interactive Media Lab Dresden,\\TU Dresden}
\and Lukas Gerlach\thanks{e-mail: lukas.gerlach@tu-dresden.de}\\ %
    \parbox{1.8in}{\scriptsize \centering Knowledge-Based Systems Group,\\TU Dresden}
\and Tobias Wieland\thanks{e-mail: tobias.wieland@mailbox.tu-dresden.de}\\ %
    \parbox{1.8in}{\scriptsize \centering Interactive Media Lab Dresden,\\TU Dresden}
\and Alex Ivliev\thanks{e-mail: alex.ivliev@tu-dresden.de}\\ %
    \parbox{1.8in}{\scriptsize \centering Knowledge-Based Systems Group,\\TU Dresden}
\and Markus Kr\"{o}tzsch\thanks{e-mail: markus.kroetzsch@tu-dresden.de}\\ %
    \parbox{1.8in}{\scriptsize \centering Knowledge-Based Systems Group,\\TU Dresden}
\and Raimund Dachselt\thanks{e-mail: dachselt@acm.org}\\ %
    \parbox{1.8in}{\scriptsize \centering Interactive Media Lab Dresden,\\TU Dresden}}

\usepackage{xspace}
\usepackage{underscore}

\newcommand{\etal}[1]{{#1}~et~al.}
\newcommand{\ie}{i.e.,\xspace}

\newcommand{\eg}{e.g.,\xspace}
\newcommand{\etc}{etc.\xspace}

\newcommand{\nemo}{\emph{Nemo}\xspace}
\newcommand{\nemoweb}{\emph{Nemo Web}\xspace}

\newcommand{\nev}{\emph{nev}\xspace}
\newcommand{\nevF}{\emph{Nemo Explain Visualizer}\xspace}

\newcommand{\qone}{\emph{T1}\xspace}
\newcommand{\qtwo}{\emph{T2}\xspace}

\teaser{
  \centering
  \includegraphics[width=\linewidth]{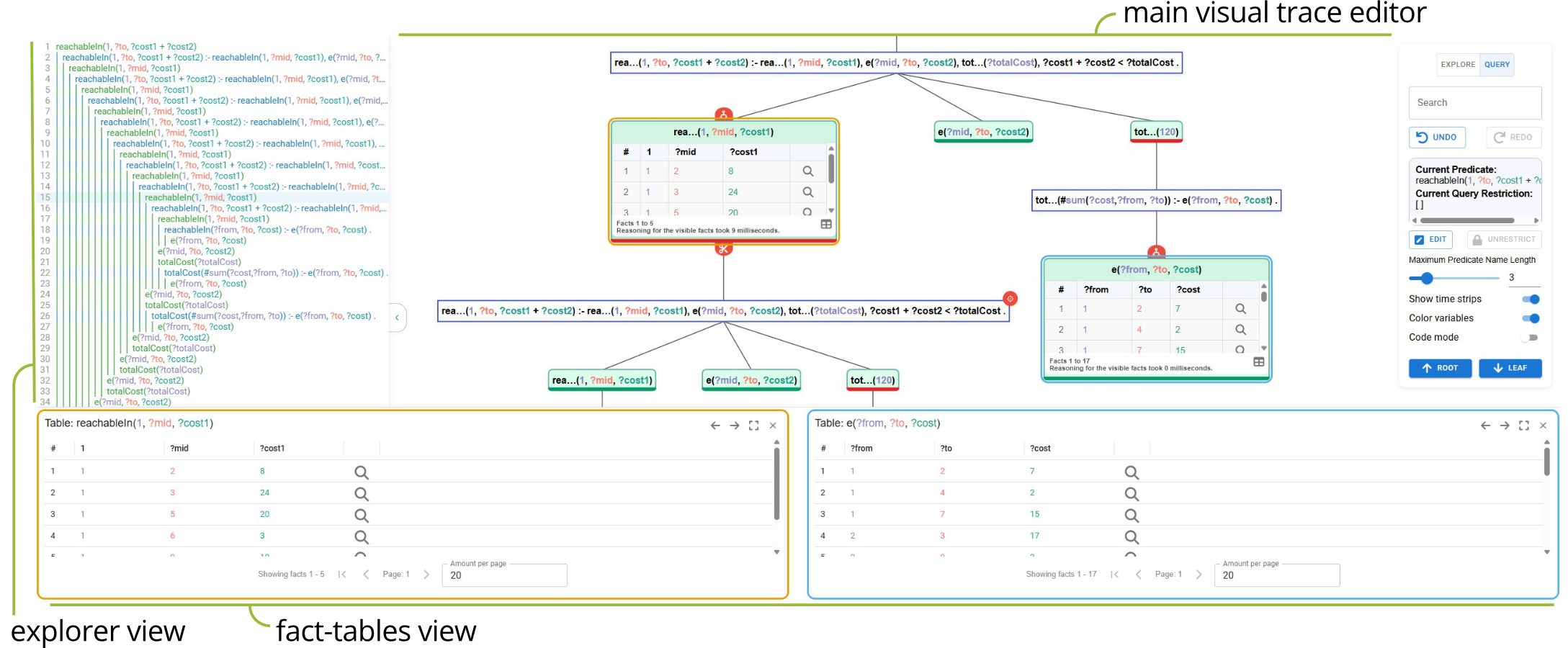}
  \caption{Screenshot of \nevF (\nev), illustrating its three main interface components: the indented tree \emph{explorer view} on the left, the \emph{main visual trace editor} on the center, and \emph{fact-tables view} at the bottom. The search bar, undo/redo, query restriction and other menu buttons are floating on the right.}
  \label{fig:teaser}
}

\abstract{Nemo Explain Visualizer (nev) is an interactive visual query tracer and builder for Nemo, a powerful Datalog reasoner with extended features. Our tools \add{were developed with and for} expert users. However, considering the lack of resources to learn Datalog and similar declarative logic programming languages, we conducted a qualitative user study to assess how our tools \add{might help students}. The study, interviewing 14 participants with varying levels of involvement with the content of a university course on knowledge graphs, revealed a very positive assessment of our tools, which strengthens the value of visual explanation tools beyond their intended use.} % end of abstract


\providecommand{\remove}[1]{\ignorespaces}

\providecommand{\add}[1]{{#1}}

\begin{document}


\firstsection{Introduction}

\maketitle

An important and diverse set of computing technologies is based on \emph{rules} -- simple \emph{if-then} statements that express logical implications --
which are combined into \emph{logic programs}. 
The declarative rule language \emph{Datalog}~\cite{Maier+:DatalogHistory18} is a core approach that recently received renewed interest for code analysis (\eg Semmle/Github), database query (\eg Google Logica), web data extraction, and more~\cite{Kroetzsch2025:Datalog}.
Being \emph{declarative} means Datalog and its relatives let users focus on 
\emph{what} should be achieved rather than on \emph{how} it should be computed.

In the early 1990s, graph visualizations of logical computations were said to increase user understanding~\cite{Dedex} and \emph{derivation trees} to ``closely mirror the manner in which users think of rule-based programs''~\cite{AroraRRSS93}.
This intuition still informs approaches to Datalog explanation and debugging \cite{DeclarativeDebuggingOfDatalogPrograms,MereMortals}.
% At the same time, modern programming environments have been adopted to provide much improved user experience for logic programmers~\cite{ASPChef,Vadacode,SeaLion,ASPIDE}.
%
However, graph visualization is largely missing from modern logic programming IDEs~\cite{ASPChef,Vadacode,SeaLion,ASPIDE}.
\emph{Souffl\'{e}} \cite{Souffle} and \nemo \cite{IGMSK2024} are modern Datalog reasoners that can compute derivation tree structures, but the only recent attempt to visualize them is a prototype we
developed for Nemo~\cite{GIMMDK2024}.
Likewise, we lack user-centered evaluations of visual tools for Datalog reasoners.
We must therefore ask whether visualizations in logic programming are as helpful and relevant as originally thought. 

\begin{figure*}[t]
  \centering
  \includegraphics[width=1\linewidth]{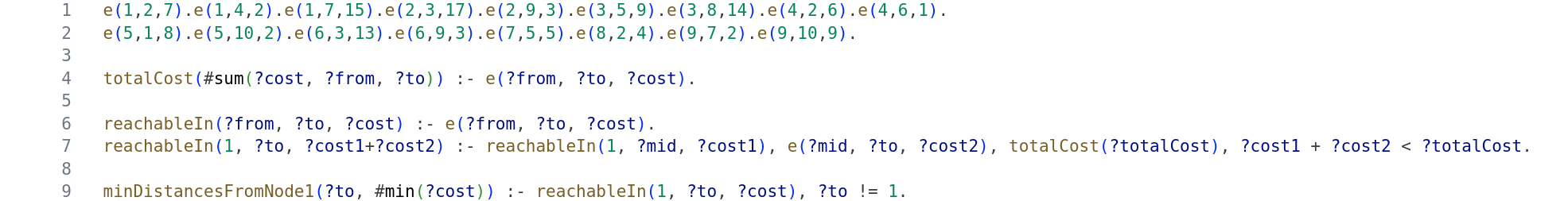}
  \caption{\label{fig:example-nemo-program} Screenshot of \nemoweb with a Datalog program that computes paths of minimal weight in a graph.} 
\end{figure*}

We tackled this question by (1) developing \nevF (\nev), a web-based visual explanation tool for Datalog reasoning, and (2) conducting a 
%first-of-its-kind 
qualitative user study \add{to gather initial evidence} of its benefits for declarative logic programming.
Our Datalog environment consists of: (1) \nemo, the Datalog reasoner; (2) the \nemoweb frontend, with runtime controls and a rich program editor; and (3) \nev, shown in~\autoref{fig:teaser}, which allows users to interactively manipulate query traces and retrieve computed facts on demand.
\nemo and \nemoweb~\cite{IGMSK2024} received significant extensions to compute
explorable explanation data on-demand, and \nev is a new system developed to enable this exploration visually. %It is inspired by relational database query execution visualizers like \emph{pev2}~\cite{PEV2}.
%Such tools visualize execution plans and provide insights about resource usage for optimization and debugging,
%all based on the different (but also graphical) execution traces of relational databases. 
All of our tools are free and open source (see \href{https://imld.de/nev}{https://imld.de/nev}).
% However, these tools are often created with experts in mind and seldom incorporate tutorials, tests, or guided walkthroughs typical of e-learning tools~\cite{10.4018/IJITWE.333638}. Dedicated learning resources like those are either general-purpose (\eg digital whiteboards) or require a considerable effort to create, distribute, and maintain, especially in niche communities like the ones of declarative logic programming languages. 
% Overall, our Datalog environment provides:
% \begin{itemize}
% \item Fully browser-based Datalog support with advanced IDE features and proof tree visualization, removing the need for installation or server capacity for widest possible access.
% \item Scalability to datasets of millions of facts, both in terms of reasoning performance (low latency, low memory consumption) and in terms of visual interaction (interactive stepwise exploration, paging, zoom \& pan, search, etc.).
% \item Support for querying explanations based on their \emph{shape} (dubbed ``type 2 query'' below), which can be used to discover computation patterns in new ways.
% \end{itemize}
% This paper focuses on the aspects of user interaction and visualization, and
% omits implementation details of \nemo. These details and scalability studies are reported by \etal{Ivliev}~\cite{IGMSK2024}.

We designed \nev\add{~in collaboration with domain \emph{experts},} for debugging and development. 
However, \add{for our evaluation, we target mostly early but also advanced Datalog \emph{learners}, to ensure a broader user group}.
The driving assumption behind this decision is that a well-designed system for programming also supports \emph{computational thinking}~\cite{computationalThinking}, which is helpful for both learners and experts. Furthermore, involving users throughout the design process is important for human-centered design~\cite{Garreta-Domingo2018}. Therefore, despite the state of our tool (\ie without specific learning features~\cite{10.4018/IJITWE.333638}), our evaluation with learners aims at making \nev accessible to our broader target group, \add{with the expert side already partially accounted for, because of our user-centered design process}. 
We therefore conducted 14 interviews to study the value of our Datalog environment for explainability \add{at lower expertise levels}.
The results of this study inform the prioritization of tool developments in nascent communities, since visual explanation tools might alleviate the need for tailored learning resources. More so, the results of our study support the intuition that visualization has significant potential in this domain, and therefore deserves renewed focus in the logic programming community.

\section{Background \& Related Work}
This section provides basic notions of declarative logic programming, related visual tools, and theory on computational thinking. 

\subsection{Declarative Logic Programming} \label{sec:rw-dlp}
Declarative logic programming languages at their core consist of simple ``if-then'' rules.
For example, the rule $B(x) \leftarrow A(x)$ means that if the fact $A(c)$ holds, then $B(c)$ is also true.
In practice, it is common to use Datalog extensions
to also support datatypes, negation, aggregation, or existentially quantified variables.
Prominent examples of Datalog systems with subsets of these features include 
\emph{RDFox}~\cite{RDFox}, 
\emph{Soufflé}~\cite{Souffle}, 
\emph{VLog}~\cite{vlog}, and 
\nemo~\cite{IGMSK2024}.
% Without going into too many technical details, the semantics of the mentioned systems still enforce some restrictions on the use of negation and aggregates. 
% Logic programming systems that lift these restrictions include 
% answer-set programming systems (\eg Clingo~\cite{clingo} or DLV~\cite{dlv}) and
% Prolog implementations (\eg XSB \cite{SwiftWarren:XSBTabling12} or \cite{swi-prolog}), which can also support Datalog reasoning.

We can trace how any fact in the result of a Datalog program came into existence by backtracking the rules that contributed to the inference, all the way to the initial facts, resulting in a proof tree. This makes the systems inherently explainable.
\emph{Soufflé} and \nemo already include features to obtain such proof trees, also called traces, for a given fact in their result. 
Consider the example Datalog program from~\autoref{fig:example-nemo-program}, written in \nemoweb. It encodes a directed weighted graph with a ternary predicate \texttt{e(from, to, weight)} and
stores the \emph{transitive closure} of the edges in the \texttt{reachableIn} predicate. 
For example, the edges $1$ to $2$ with weight $7$ and $2$ to $9$ with weight $3$, imply
that $9$ is reachable from $1$ with a combined weight of $10$, stored as \texttt{reachableIn(1,9,10)} \add{(line 7 in~\autoref{fig:example-nemo-program})}.
%Finally, the program computes for each node $n \neq 1$ the minimal weight to get from $1$ to $n$ stored in the predicate \texttt{minDistanceFromNode1}.
Tracing this derived fact, shows us its proof tree. In this case, we expect a tree that has \texttt{reachableIn(1,9,10)} at its root, with \texttt{edge(1,2,7)} and \texttt{edge(2,9,3)} as \add{leaves}.

\subsection{Visual Tools to Explain, Trace, and Inspect Queries} \label{sec:rw-vis}
Visualization has been widely employed in the field of knowledge representation and reasoning, \eg for ontology analysis and editing~\cite{WEB_PROTEGE, crowd, UnSHACLed} tasks. 
% A great source of examples for this is the collection of visualization tools for Wikidata (see \href{https://www.wikidata.org/wiki/Wikidata:Tools/Visualize_data}{wikidata.org/wiki/Wikidata:Tools/Visualize\_data}). 
In areas closer related to logic programming, IDEs provide basic visualization features such as syntax highlighting~\cite{SeaLion,ASPIDE,Vadacode}
and tools like ASP Chef~\cite{ASPChef} even allow to build complex ASP programs by dragging and dropping basic building blocks. This also includes visualization capabilities for common structures, \eg graphs, that are encoded in the program. 

We refer to visualization tools concerned with explaining the outcomes of reasoning processes as \textit{visual explanation tools}. 
Some examples include proof visualizers~\cite{
Logichart,
AroraRRSS93,
MAKLBDCGF2023,
Dedex} and database execution plan visualizers~\cite{
pgmustard, 
PEV2, 
picassoDBVisualizer,
mysqlvisualexplain}. 
The former focus on presentation and interactivity of the proofs, while the latter support debugging slow queries. 
The proofs/traces supported by these systems are often trees or directed acyclic graphs (DAGs), which may contain repeating patterns and recursive structures. 
Naturally, these tools employ techniques for visually exploring node-link diagrams~\cite{VIS_LINK}, such as multiple coordinated views~\cite{COORDINATED_MULTIPLE_VIEWS}, and details on demand within nodes and onto juxtaposed views~\cite{compositeVis}. 
The survey by \etal{Dudáš} lists components typical to ontology visualization tools, such as radar views (overview), history (in case of editing), graphical zoom and pan, entity focus, and search and highlight~\cite{ontologyvis}. Therefore, we used it to inform several features of \nev. 

% \nemo originally incorporated a proof visualizer to illustrate its execution traces~\cite{GIMMDK2024}, embedded into a modal dialog in \nemoweb, but did not show the facts computed by each rule. A more suitable approach would show these facts and additional performance information (as done in execution plan visualizers), all while allowing users to distribute the interface across multiple windows, screens or even devices (resembling development environments with multiple tabs or applications). 

% \begin{figure*}[t]
%     \centering
%     \includegraphics[width=.9\linewidth]{figures/paths-of-length-three.png}
%     \caption{\label{fig:paths-of-length-three} Screenshot of \nev's main trace view with a \qtwo query and its result, which corresponds to the same trace populated with retrieved facts. This \qtwo queries for paths of exactly three edges from the code on~\autoref{fig:example-nemo-program}. Green nodes correspond to predicates (and can be expanded to show the facts as tables inside), while the blue-bordered nodes with sharp corners correspond to rules.} 
% \end{figure*}
\subsection{Evaluating Visual Explanation Tools for Learning}\label{sec:rw-eval}

We based the design of our evaluation on the concept of \textit{computational thinking} (CT)~\cite{computationalThinking}, which refers to the employment of computing strategies such as decomposition and abstraction to tackle complex problems. Although CT has been discussed with various terminologies~\cite{CTterminology1, CTterminology3} it is roughly described as a mixture of skills of data representation, algorithmic design, and pattern recognition. 
Since the subject of our study is declarative logic programming, it follows that a tool that supports CT in this computer science domain should support not only working on it, but also learning it. 
Furthermore, for human-centered design, users should be involved in early evaluations of the entire user experience (UX)~\cite{Garreta-Domingo2018}. We therefore evaluate CT and UX for our programming environment, as these assessments correspond to learners and experts simultaneously.

\section{Nemo Explain Visualizer (nev)}
The name and design of \nev are heavily inspired by Postgres Explain Visualizer 2 (\emph{pev2})~\cite{PEV2}. In this section, we explain our extension of this concept through the visual construction and manipulation of shape-based queries.

\subsection{\nemo and the Tree-Shape Query}

For a given fact, \nemo can compute a proof tree (trace) to witness how the fact came into existence during the reasoning process. We refer to these queries as type 1 (\qone). 
The new type of query (\qtwo) instead receives the \emph{shape of a proof tree} and populates it with facts from all traces that include a partial proof tree of the given shape (if any exist).
\qone is useful for explaining individual facts, but \qtwo was introduced mainly for debugging purposes to see how different rule applications interact without the need of changing the program.
On the example from~\autoref{fig:example-nemo-program},
to know which facts of \texttt{reachableIn} correspond to paths with exactly three graph edges, one can build a \qtwo with a tree shape of two applications of the rule in line 7 after one application of the rule in line 6. \nemo then returns the input shape with facts from trees matching this shape. In this case, the desired facts can be read on the root node. %In other words, \qtwo{}s use a tree shape as input and output, and populate the nodes within the tree shape with facts, if any can be retrieved. 
\qtwo also includes execution times of applications to assist in analyzing performance. \add{\qtwo{}s are formally described for Nemo, and their performance measured, in~\cite{DGHIKMM2026}.}

\subsection{Visual Analysis and Query Editing}\label{sec:visualization}
\newcommand{\dudas}[1]{{#1}}
We developed \nev to simultaneously use the graphical trace for analysis and query editing, thanks to \qtwo{}s using \nemo traces as input and output. 
\add{Upon running a program on Nemo Web, the user is presented with tables that contain the derived facts for all rules. Choosing a derived fact opens \nev with the respective \qone{} in a new browser tab.}
The interface of \nev consists of a \emph{main visual trace editor}, an indented tree \emph{explorer view} and \emph{fact-tables view}, as shown in~\autoref{fig:teaser}. 
We distinguish two types of nodes: \textit{predicates} using green highlights (\ie border and fill color) and \textit{rules} using blue borders. 
One can \dudas{zoom} and \dudas{pan} the main view for navigation, but hovering or clicking on the indented tree highlights and brings nodes to the center of the main view for faster navigation. Clicking on a predicate node in the main view reveals the facts associated with it on a small table embedded in the node. These tables can be pinned to the fact-tables panel at the bottom for inspection (using pagination and counts) as well as comparison \eg for debugging purposes. Opposed to the modal dialog approach previously available to \nemoweb~\cite{GIMMDK2024}, \nev is loaded onto a separate browser tab. The user can thus configure their screen(s) flexibly (\eg side-by-side, different monitors) without cluttering the code editor view. 

\qone{}s establish a \textit{query restriction} that specifies the facts to look for. Without it or upon manipulating the tree shape (\ie adding or removing nodes), \qtwo{}s are created.
Users may \textit{add} or \textit{remove rules} using the green and red buttons above and below predicates nodes where this is viable (\ie \qtwo{}s include potential rules to add). When more than one rule can be added, the choices are shown via dialogs. Users can also \textit{focus on a rule} by clicking the target icon on rule nodes. This constructs a \qtwo of a single rule and its surrounding predicates, which \add{is} useful for inspecting all applications of a rule across all possible derivations. 
To support back and forth between the different traces corresponding to these features, we incorporated undo and redo features that save the tree (including restrictions) after any modification. 
To additionally mitigate the risk of accidental clicks, as well as to support \dudas{incremental exploration}, we also enable an \textit{exploration mode} that disables editing, where users can collapse and expand the tree from any node, and focus on specific rules by blurring the rest of the tree instead of removing it. We also provide search and name shortening options.

\section{Evaluation}

We conducted a qualitative study of our \add{tools} on students with some level of involvement in declarative logic programming. 
As indicated in~\ref{sec:rw-eval}, this is motivated by the close relation between our use case and computational thinking (CT)~\cite{computationalThinking} and the need to involve users in early design phases~\cite{Garreta-Domingo2018}. 
We conducted semi-structured interviews, following a think-aloud protocol. 
The research questions of this experiment~\cite{osfNemoNev} were: 
\textbf{RQ1:} How do the interactive editor features and visual tracing help students understand rule-based reasoning? 
\textbf{RQ2:} How does subject literacy (\ie on declarative logic programming) influence the use of \nemo+ \nev?
\textbf{RQ3:} How does the user experience and task difficulty scale with \nemo+ \nev with respect to task complexity? 
All research questions serve to validate \nev, but RQ1 and RQ3 additionally address CT.
The study was preregistered~\cite{osfNemoNev} and our supplementary material includes all used forms, documents, and anonymized data and results. The version of \nemoweb and \nev used for this study is available online at \href{https://tools.iccl.inf.tu-dresden.de/nemo/nev-study-2025}{tools.iccl.inf.tu-dresden.de/nemo/nev-study-2025}. 

% \begin{figure}[!t]
%   \centering
%   \includegraphics[width=1\linewidth]{figures/expertise.pdf}
%   \caption{\label{fig:expertise} Counts of self-assessed expertise from the participants from None (lowest) to Expert (highest) in the topics of Datalog, \nemo, querying tools (QT), and data visualization (DV) tools.} 
% \end{figure}

\paragraph*{Participants:} We interviewed 14 participants (3 female, 11 male) enrolled in bachelor (3), master (4), or diploma (4) study programs. Regarding expertise, we used a scale from 0 (no experience) to 5 (expert). Only two participants scored themselves as Experts in \nemo and Advanced in Datalog. Every other score, including by the same participants in the other topics, was 3 (Intermediate) or below, with 4 participants averaging 1 or below (very low to no experience overall). Participation was compensated with 20 EUR. 

\begin{figure}[!t]
  \centering
  \includegraphics[width=1\linewidth]{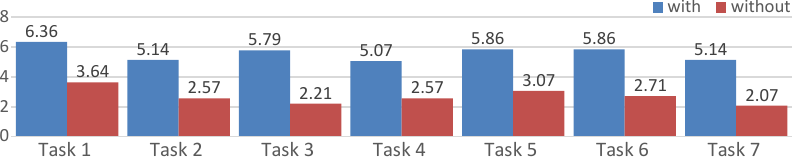}
  \caption{\label{fig:seq} Average SEQ assessments for all tasks (1 = very difficult, 7 = very easy) \add{for the \emph{perceived} task difficulty using our tools (blue) and the \emph{imagined} difficulty without our tools (red). Due to the absence of a baseline, these values support our tools only qualitatively.}} 
\end{figure}

\begin{figure*}[tbp]
  \centering
  \includegraphics[width=1\linewidth]{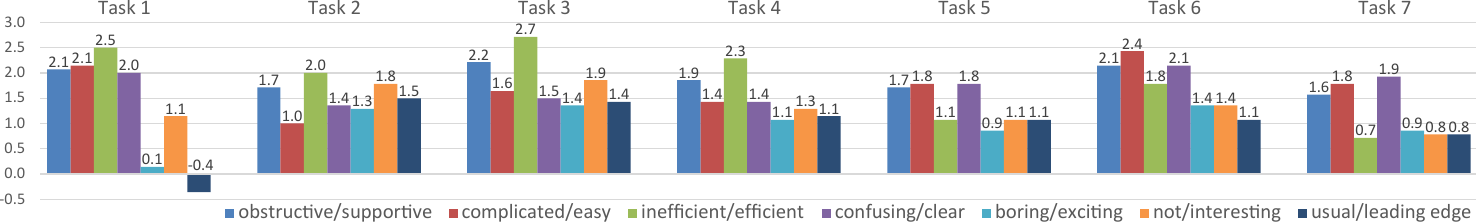}
  \caption{\label{fig:ueq-s} Average UX assessments for all tasks. UEQ-S items are normalized from -3 (horribly bad) to +3 (extremely good). Except for usual/leading edge for Task 1, all assessments are positive, in particular the pragmatic ones (the first 4 from left to right).}
\end{figure*}

\paragraph*{Apparatus:} The interviews were conducted in person on a laboratory setting with two interviewers per participant, one who guided the participants through the tasks, the other for taking notes but also intervening in case of questions. The participants would use one of two desktop computers with a 4K UHD 32" monitor, Intel i7 processors (-11700K and -13700K), and NVIDIA GeForce RTX 3080 and 4090, on the Brave browser on Windows 11. 

\paragraph*{Procedure:} We collected informed consent, demographic data, audio, and screen captures. The experiment took in average 77 minutes, with a time limit of two hours that no participant exceeded. 
We devised a scenario where a faulty Datalog program must be iteratively corrected and improved until it corresponded to the program shown in \autoref{fig:example-nemo-program}. 
This is an ecologically valid scenario common when developers leave behind pieces of code that no other employee maintained or looked at. 
The participants would work with this program over seven tasks, always in the same order, and could freely converse with the interviewers. 
\add{The interviewers did not assist the participants in completing the tasks, but provided general hints about \eg syntax and how Datalog rules work, upon request.} 
Each task would be assessed in terms of task difficulty (TD) and user experience (UX). 
After the interviews were finished, we provided space for final general comments about our tools.

\paragraph*{Tasks:} The given program should compute the minimal cost between nodes in a weighted graph (10 nodes, 17 edges), but this information was withheld. The code initially contained syntactic errors, undefined variables, \etc. 
All participants would solve the following 7 tasks in order: 1) correct the breaking errors in the code; 2) inspect various traces to understand the program and rename symbols accordingly; 3) find paths of minimal cost and shorter paths (regardless of cost) using \nev (without editing the program further); 4) inspect and correct a performance issue in the code; 5) modify the \nev queries to find paths of specific length; and finally, assess the scalability of our tools by finding lowest cost paths on 6) larger (100 nodes, 444 edges), and 7) much larger graphs (500 nodes, 2284 edges). For the first task, only \nemoweb would be used, but for the other six, the participants could use both \nev and \nemoweb freely.  
These tasks correspond to the CT principles and relation to user experience discussed in~\ref{sec:rw-eval}. Namely, identifying a graph data structure (tasks 1 and 2), modifying a recursive algorithm (1 and 4) and re-purposing features and interface components (3, 5). Tasks 6 and 7 directly address RQ3. 

\paragraph*{Data Collection:}  We used OBS Studio~\cite{obs} for screen and audio capture. For subtitle generation/transcription we used Buzz~\cite{buzz} with Whisper~\cite{whisper}. Task difficulty (TD) was assessed using Single Ease Question (SEQ)~\cite{DBLP:conf/chi/SauroD09} scales, while for user experience (UX) we used a modified version of the Short User Experience Questionnaire (UEQ-S)~\cite{ueq-s}. SEQs are 7-point Likert scales from \textit{very difficult} to \textit{very easy}, while UEQ-S is composed of eight 7-point Likert scales. For TD, we used two SEQs per task and asked the participants to rate the difficulty of solving the tasks using our tools (\nemoweb for task 1, both \nemoweb and \nev for the other tasks) \add{and the imagined} difficulty of completing the tasks without our tools. For UX, we excluded the 7th question of the UEQ-S due to its similarity to the 8th. This helped us maintain a reasonable interview time but precludes statistical conclusions about the hedonic nature of our tools. Nevertheless, the 7 questions remain meaningful descriptors of our tools. We took notes of the participant's comments, bugs encountered, and suggestions for features. 

\paragraph*{Results \& Discussion:} 
All participants completed all the tasks within the time limit.  \autoref{fig:seq} shows the average TD \add{scores}, while \autoref{fig:ueq-s} shows the equivalent for UX. Our tools were very positively received: \eg ``It was pretty intuitive, very nicely done'', ``It is helpful to have this tree with the derivations, and adding and removing rules, also restricting and unrestricting''. 
Regarding TD, the SEQ scores show a very positive reception of our tools, which is unsurprising considering the lack of alternatives for declarative logic programming (\eg text editors and command-line interfaces). When switching from just \nemoweb (task 1) to also using \nev (tasks 2 and on), we received comments like ``before using \nev I would say this was moderately hard and after using \nev, I think more easy''. 
All averages (using our tools versus not using them) differ by at least 2.5 points, with the largest differences on tasks 3 (3.57), 6 (3.14), and 7 (3.07). These tasks correspond to the scalability comparison of RQ3, and a reasonable reduction of the difference can be noted as the graph size increases. 
%Interestingly, the score difference between tasks 3 and 6 is larger than between 6 and 7, despite the disproportionate growth of the graph. This might be related to the participants becoming more used to \nev after tasks 4 and 5. However, 
We interpret these results as a positive indication of the scalability of our approach, despite ample room for further features to improve scalability. This interpretation is supported by the similarly high UX scores and the matching trend between tasks 3, 6, and 7. However, a mixed sentiment was expressed by some participants, \eg ``\nev is probably not made to read a path along such a long derivation, but even then it works well'' and ``It doesn't scale very well. Still much better than just using \nemo'' (referring to \nev). 
Towards CT, these positive results are influenced by the interviewer assistance provided to alleviate the learning curve and limit the duration of the interviews, \eg ``Without any guidance it would be much more complicated to figure out, but I guess that's what tutorials are for''. Even when acknowledging this, our participants showed positive sentiments, though: ``once I understood what the buttons do, I can easily perform any task''. 
Regarding UX, the UEQ-S questions were also overall very positive, with a single exception on Task 1. Since this task only introduced \nemoweb and its language server features that are common to many (more popular) programming languages, this is understandable. A participant indicated: ``It follows what I expect an editor to look like (...) it's also boring, which I consider to be a good thing''. In fact, the same task also received the second highest average score (for efficiency, highest on task 3), which was also emphasized by several participants, \eg ``I like the snappiness''. With respect to RQ2, the participants with higher expertise had an easier time both completing the tasks and using the tools. This was reflected by their understanding of the program before completing Task 2. Namely, the most experienced participant
identified the graph structure and purpose of the program during Task 1. 
Those with less experience would only understand the program (after hints by the interviewers) shortly before task 3. However, most participants were able to correctly answer validation questions asked throughout the interview, such as identifying equivalent ways to unrestrict the queries. 
Altogether, the assessments to TD and UX address RQ1, showing that our language server, visual tracing, query building, and performance inspection features are well-received for working with and learning declarative logic programs. 
Furthermore, the higher UX scores in pragmatic over hedonic qualities reflect the prioritization of feature developments from our side. 
%Given the positive reception of the tools despite this imbalance, we are inclined to believe that the pragmatic qualities should take precedence when it comes to explainability and learning. 
%One related factor may be trustworthiness, as overly-embellished tools may evoke distrust. 
During the interviews, several bugs were identified by the participants, such as on the queuing of requests when multiple pagination requests are triggered in quick succession, and with the replace symbol interaction on the code editor. 
%While, in the name of reproducibility, we provide access to the version of our tools used for this study on \href{https://tools.iccl.inf.tu-dresden.de/nemo/nev-study-2025}{tools.iccl.inf.tu-dresden.de/nemo/nev-study-2025}, 
\add{These issues have been addressed in the respective repositories 
%of our tools (Nemo, Nemo Web and \nev), 
and several post-study updates are already reflected in the screenshot of~\autoref{fig:teaser}. The improvements include variable coloring, colored strips (green to red) to assess performance per node at a glance, and the ability to jump to and highlight lines in the code on the linked Nemo Web tab by clicking on the corresponding rules or focus buttons}.

\add{\paragraph*{Limitations:} The absence of a baseline for comparison (\eg a command-line interface) precludes quantitative observations regarding the TD scores, which should therefore only be taken as evidence of subjective support for our tools. Furthermore, future evaluations of our tools should invite more participants that sample the entire target group (\ie experts and students alike), gather feedback from educators, and quantitatively analyze learning benefits (\eg measured comprehension, grades) in an unsupervised setting (\ie without interviewer guidance). 
Furthermore, scalability has only been superficially discussed thus far. Theoretically, much larger traces than the ones in our study may be constructed. Thus, we aim to explore node grouping, focus+context techniques, and \eg off-screen indicators of (infinite) recursion in future versions.}

\section{Conclusion}

We presented \nev and the results of an evaluation of our declarative logic programming environment. By designing our study in an incremental way around tasks that require understanding of algorithms and data structures, we also assessed the potential of our tools to support computational thinking. We received very positive assessments of our tools with respect to task difficulty and user experience. %, which validates our contributions on their own and for educational scenarios. 
%Furthermore, the thoughtful conversations with our study participants also revealed a number of opportunities for improvement, such as collapsing repeating patterns in the traces, embedded heatmap encoding performance measures directly in the code, and testing for recursive rules. 
Our work illustrates the value that visual explanation tools can provide for communities with scarce learning resources \add{instead of using general-purpose learning tools}. %as opposed to the application of general-purpose learning tools to arbitrary domains. 

\acknowledgments{
\add{The authors thank Afshin Zanganeh for his contributions to nev. No AI was used in the creation of this paper or its contributions.} 
This work is funded 
by Deutsche Forschungsgemeinschaft (DFG) 
under Germany’s Excellence Strategy: EXC 2050/2, 390696704 – “Centre for Tactile Internet” (CeTI);
by DFG grant 389792660 as part of \href{https://perspicuous-computing.science}{TRR~248 -- CPEC};
%by Bundesministerium für Bildung und Forschung (BMBF) under European ITEA project 01IS21084 (\href{https://www.innosale.eu/}{InnoSale});
by Bundesministerium für Bildung und Forschung (BMBF) and Saxon State Ministry for Science, Culture and Tourism (SMWK) in \href{https://www.scads.de}{Center for Scalable Data Analytics and Artificial Intelligence} (ScaDS.AI, SCADS22B);
and by BMBF and German Academic Exchange Service (DAAD) in project 57616814 (\href{https://secai.org/}{SECAI}, \href{https://secai.org/}{School of Embedded and Composite AI}).}

\bibliographystyle{abbrv-doi}

\bibliography{template}
\end{document}